# A Simplified Dynamical Model for Tuned Wireless Power Transfer Systems

Hongchang Li, *Member, IEEE*, Jingyang Fang, *Member, IEEE*, and Yi Tang, *Senior Member, IEEE*

*Abstract* – Dynamical models of wireless power transfer (WPT) systems are of primary importance for the dynamical behavior studies and controller design. However, the existing dynamical models usually suffer from high orders and complicated forms due to the complex nature of the coupled resonances and switched-mode power converters in WPT systems. This letter finds that a well-tuned WPT system can be accurately described by a much simpler dynamical model. Specifically, at the tuned condition, the existing dynamical model can be decomposed into two parts. One is controllable and the other one is uncontrollable. The former should be considered in the modeling while the latter can be ignored because it always exponentially converges to zero. For illustration, the recently proposed zero-voltage-switching full-bridge pulse-density modulation WPT system is modeled as an example since such a system can efficiently operate at the tuned condition with soft switching and control capabilities. The derived model was verified in experiments by time-domain and frequency-domain responses.[1]

*Index Terms* – Dynamical model, wireless power transfer.

## I. INTRODUCTION

Wireless power transfer (WPT) systems are becoming widely used in electric vehicles (EVs), consumer electronics, medical devices, factory automations, etc. [2]. Some of these applications suffer from fast parameter changes and require rapid control and protection, e.g. the road-powered EVs [3]. Consequently, the dynamical behaviors of WPT systems need to be well studied and the analytical dynamical models are of primary importance.

The most common dynamical modeling methods for WPT systems include the generalized state-space averaging (GSSA) [4], extended describing functions (EDF) [5], and dynamic phasors [6]. However, the dynamical models derived using these methods suffer from high orders and complicated forms due to the complex nature of the coupled resonances and switched-mode power converters in WPT systems. For example, a simple series-series compensated WPT system shown in Fig. 1, which has 4 resonant elements and 1 filter capacitor, has to be modeled by a 9[th]-order model [5]. To reduce the model order, previous studies proposed model order reduction techniques and presented 5[th]-order models for the system shown in Fig. 1 [7]. Furthermore, the experimental results given in [7] imply that the system may be described by an even simpler model, and drive a further study on the dynamical modeling.

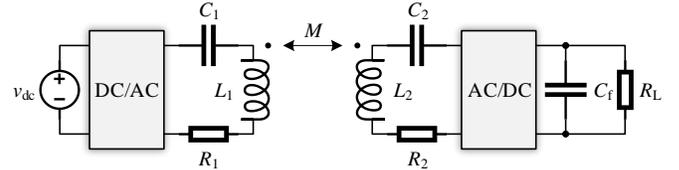

Fig. 1. A series-series compensated WPT system.

This letter finds that when the system is well tuned, the 5[th]-order real-valued dynamic phasor model in [7] can be decomposed into two parts. One is controllable and the other one is uncontrollable. The former includes the real part of the transmitter resonant current phasor, the imaginary part of the receiver resonant current phasor, and the dc output voltage. The latter includes the imaginary part of the transmitter resonant current phasor and the real part of the receiver resonant current phasor. The controllable part can be used as a simplified dynamical model for tuned WPT systems.

## II. TUNED WPT SYSTEM

It is well known that tuned resonances are essential for loosely coupled near-field WPT systems. Many studies in the literature take the tuned condition as a basic assumption of their analysis. However, for the purpose of soft switching, WPT systems are usually detuned to make an inductive or capacitive load for the full-bridge or half-bridge inverter [8]. Moreover, the variable frequency control method suggested by Qi [9] inevitably detunes the system. Fortunately, the recently proposed zero-voltage-switching (ZVS) full-bridge pulse-density modulation (PDM) WPT system can well satisfy the tuned condition while maintaining soft switching and control capabilities [10]. With the improved PDM strategy [11], such a system has become one of the best platforms for the dynamical behavior studies.

Fig. 2 shows the circuit diagram of a ZVS full-bridge PDM WPT system. The system consists of a dc voltage source, a ZVS full-bridge inverter, a pair of coupled resonators with inductances $L_1$ and $L_2$, mutual inductance $M$, capacitances $C_1$ and $C_2$, and equivalent series resistances (ESRs) $R_1$ and $R_2$, a ZVS full-bridge rectifier, a filter capacitor $C_f$, and a load resistor $R_L$. As shown in Fig. 2, the instantaneous dc side and ac side voltages and currents of the inverter are denoted by $v_1(t)$, $i_1(t)$, $u_1(t)$, and $i_{L1}(t)$, respectively. Symmetrically, the voltages and currents of the rectifier are denoted by $v_2(t)$, $i_2(t)$, $u_2(t)$, and $i_{L2}(t)$, respectively. In this system, the inverter and the rectifier are modulated as per the specified pulse densities $d_1(t)$ and $d_2(t)$, respectively, to control the conversion ratios [10, 11]. The ZVS inductors $L_{ZVS1}$ and $L_{ZVS2}$ provide ZVS currents to discharge the switch output capacitances to ensure the soft switching at various operating points.

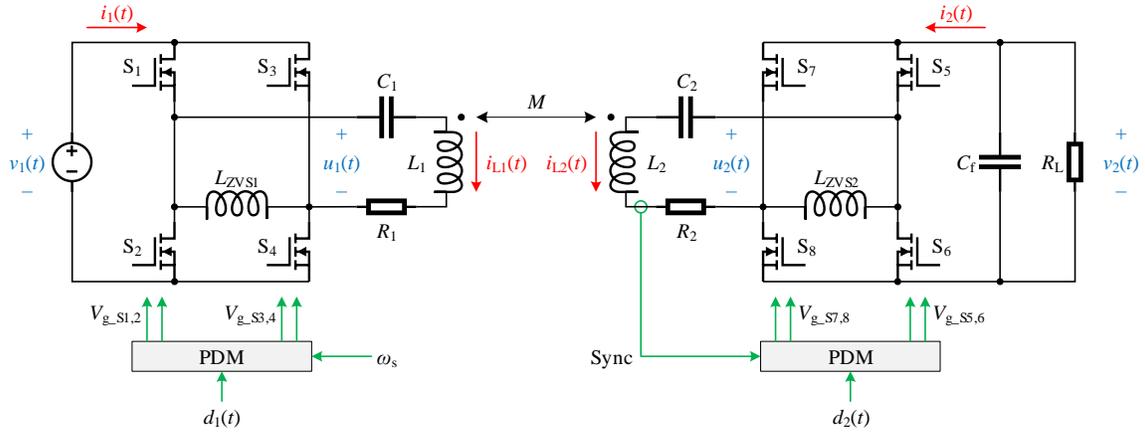

Fig. 2. Circuit diagram of a ZVS full-bridge PDM WPT system.

The system is said to be tuned when $u_2(t)$ is synchronized to $i_{L2}(t)$ with a 180° phase difference (by a zero crossing detector-based phase lock loop), and the resonant frequencies on the two sides: $\omega_{r1} = 1/\sqrt{L_1 C_1}$ and $\omega_{r2} = 1/\sqrt{L_2 C_2}$ both equal the inverter fundamental switching frequency $\omega_s = 2\pi f_s$. Under this condition, the ac equivalent input impedance of the rectifier is resistive, the phase difference between $i_{L1}(t)$ and $i_{L2}(t)$ is 90°, and the inverter has a resistive load. With these features, the dynamical behavior of the system can be described by a simplified dynamical model.

## III. SIMPLIFIED DYNAMICAL MODEL

The 5$^{\text{th}}$-order real-valued dynamical model of the system shown in Fig. 2 can be derived using the method in [7] and expressed as

$$\begin{cases} \dfrac{dI_{L1r}(t)}{dt} = \Delta\omega_1 I_{L1i}(t) - \dfrac{R_1}{L_{\omega1}} I_{L1r}(t) + \dfrac{\omega_s M}{L_{\omega1}} I_{L2i}(t) + \dfrac{S_{1r}(t)}{L_{\omega1}} V_1(t) \\ \dfrac{dI_{L1i}(t)}{dt} = -\Delta\omega_1 I_{L1r}(t) - \dfrac{R_1}{L_{\omega1}} I_{L1i}(t) - \dfrac{\omega_s M}{L_{\omega1}} I_{L2r}(t) + \dfrac{S_{1i}(t)}{L_{\omega1}} V_1(t) \\ \dfrac{dI_{L2r}(t)}{dt} = \Delta\omega_2 I_{L2i}(t) - \dfrac{R_2}{L_{\omega2}} I_{L2r}(t) + \dfrac{\omega_s M}{L_{\omega2}} I_{L1i}(t) + \dfrac{S_{2r}(t)}{L_{\omega2}} V_2(t) \\ \dfrac{dI_{L2i}(t)}{dt} = -\Delta\omega_2 I_{L2r}(t) - \dfrac{R_2}{L_{\omega2}} I_{L2i}(t) - \dfrac{\omega_s M}{L_{\omega2}} I_{L1r}(t) + \dfrac{S_{2i}(t)}{L_{\omega2}} V_2(t) \\ \dfrac{dV_2(t)}{dt} = -\dfrac{S_{2r}(t) I_{L2r}(t) + S_{2i}(t) I_{L2i}(t)}{C_f} - \dfrac{V_2(t)}{R_L C_f} \end{cases}$$
(1)

where $I_{L1r}(t)$, $I_{L1i}(t)$ and $I_{L2r}(t)$, $I_{L2i}(t)$ are the real and imaginary parts of $I_{L1}(t)$ and $I_{L2}(t)$, which are the dynamic phasors of $i_{L1}(t)$ and $i_{L2}(t)$, respectively; $V_1(t)$ and $V_2(t)$ are the moving averages of $v_1(t)$ and $v_2(t)$, respectively; $S_{1r}(t)$, $S_{1i}(t)$ and $S_{2r}(t)$, $S_{2i}(t)$ are the real and imaginary parts of $S_1(t)$ and $S_2(t)$, which are the complex-valued conversion ratios of the inverter and rectifier, respectively [7]; $\Delta\omega_1$ and $\Delta\omega_2$ are the beat frequencies given by

$$\Delta\omega_1 = \omega_s - \omega_{r1} \text{ and } \Delta\omega_2 = \omega_s - \omega_{r2} \qquad (2)$$

$L_{\omega1}$ and $L_{\omega2}$ are the equivalent inductances given by

$$L_{\omega1} = \dfrac{\omega_s + \omega_{r1}}{\omega_s} L_1 \text{ and } L_{\omega2} = \dfrac{\omega_s + \omega_{r2}}{\omega_s} L_2. \qquad (3)$$

Under the tuned condition, $S_1(t)$ and $S_2(t)$ depend only on the pulse densities $d_1(t)$ and $d_2(t)$, and can be written as

$$S_1(t) = \dfrac{2\sqrt{2}}{\pi} d_1(t) \text{ and } S_2(t) = j\dfrac{2\sqrt{2}}{\pi} d_2(t). \qquad (4)$$

Furthermore, $\omega_{r1} = \omega_{r2} = \omega_s$ yields

$$\Delta\omega_1 = \Delta\omega_2 = 0 \qquad (5)$$

and

$$L_{\omega1} = 2L_1 \text{ and } L_{\omega2} = 2L_2. \qquad (6)$$

With (4)-(6), (1) becomes

$$\begin{cases} \dfrac{dI_{L1r}(t)}{dt} = -\dfrac{R_1}{2L_1} I_{L1r}(t) + \dfrac{\omega_s M}{2L_1} I_{L2i}(t) + \dfrac{\sqrt{2}}{\pi L_1} d_1(t) V_1(t) \\ \dfrac{dI_{L1i}(t)}{dt} = -\dfrac{R_1}{2L_1} I_{L1i}(t) - \dfrac{\omega_s M}{2L_1} I_{L2r}(t) \\ \dfrac{dI_{L2r}(t)}{dt} = -\dfrac{R_2}{2L_2} I_{L2r}(t) + \dfrac{\omega_s M}{2L_2} I_{L1i}(t) \\ \dfrac{dI_{L2i}(t)}{dt} = -\dfrac{R_2}{2L_2} I_{L2i}(t) - \dfrac{\omega_s M}{2L_2} I_{L1r}(t) + \dfrac{\sqrt{2}}{\pi L_2} d_2(t) V_2(t) \\ \dfrac{dV_2(t)}{dt} = -\dfrac{2\sqrt{2}}{\pi C_f} d_2(t) I_{L2i}(t) - \dfrac{1}{R_L C_f} V_2(t) \end{cases}$$
(7)

In (7), only the 1$^{\text{st}}$, 4$^{\text{th}}$, and 5$^{\text{th}}$ equations include the control inputs $d_1(t)$ and $d_2(t)$. Besides, these three equations are all independent to $I_{L1i}(t)$ and $I_{L2r}(t)$, which are included by the 2$^{\text{nd}}$ and 3$^{\text{rd}}$ equations. Consequently, (7) can be decomposed into a controllable part:

$$\begin{cases} \dfrac{dI_{L1r}(t)}{dt} = -\dfrac{R_1}{2L_1} I_{L1r}(t) + \dfrac{\omega_s M}{2L_1} I_{L2i}(t) + \dfrac{\sqrt{2}}{\pi L_1} d_1(t) V_1(t) \\ \dfrac{dI_{L2i}(t)}{dt} = -\dfrac{R_2}{2L_2} I_{L2i}(t) - \dfrac{\omega_s M}{2L_2} I_{L1r}(t) + \dfrac{\sqrt{2}}{\pi L_2} d_2(t) V_2(t) \\ \dfrac{dV_2(t)}{dt} = -\dfrac{2\sqrt{2}}{\pi C_f} d_2(t) I_{L2i}(t) - \dfrac{1}{R_L C_f} V_2(t) \end{cases}$$
(8)

and an uncontrollable part:

$$\begin{cases} \dfrac{dI_{L1i}(t)}{dt} = -\dfrac{R_1}{2L_1} I_{L1i}(t) - \dfrac{\omega_s M}{2L_1} I_{L2r}(t) \\ \dfrac{dI_{L2r}(t)}{dt} = -\dfrac{R_2}{2L_2} I_{L2r}(t) + \dfrac{\omega_s M}{2L_2} I_{L1i}(t) \end{cases}$$
(9)

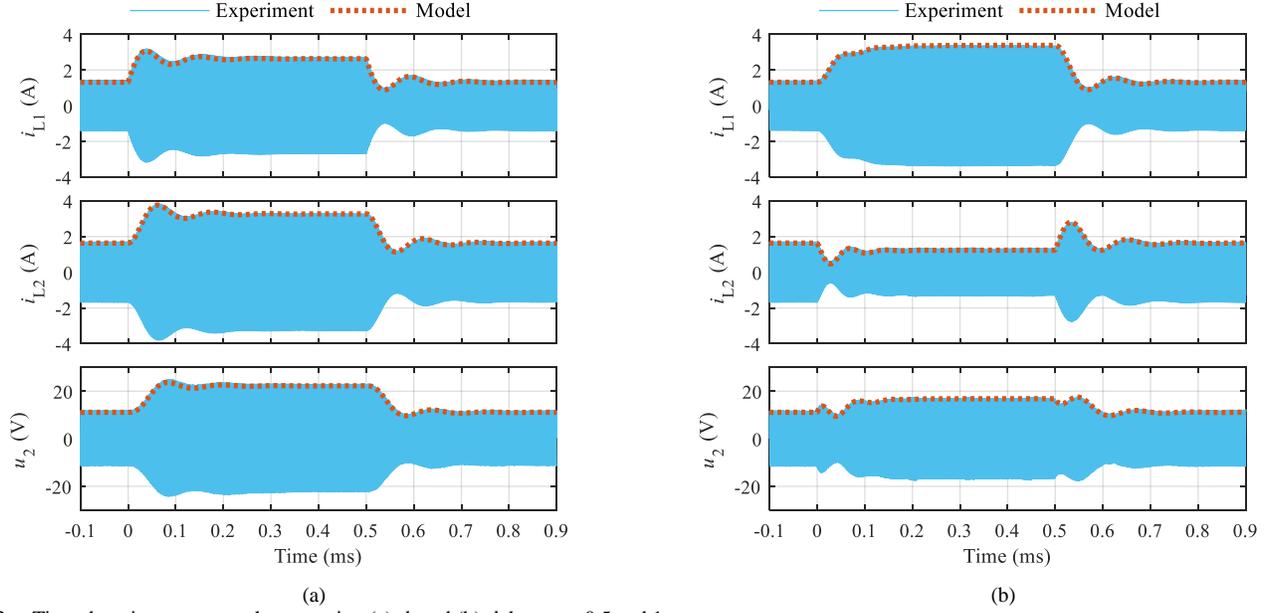

Fig. 3. Time-domain responses when stepping (a) $d_1$ and (b) $d_2$ between 0.5 and 1.

TABLE I
EXPERIMENTAL SYSTEM PARAMETERS

| Symbol | Quantity | Value |
|---|---|---|
| $\omega_s$ | Fundamental switching frequency | 5.76 Mrad/s |
| $L_{1,2}$ | Resonant inductances | 75.2 µH |
| $C_{1,2}$ | Resonant capacitances | 400 pF |
| $R_{1,2}$ | Resonant ESRs | 1.1 Ω |
| $\omega_{r1,2}$ | Angular resonant frequencies | 5.76 Mrad/s |
| $M$ | Mutual inductance | 1.17 µH |
| $C_f$ | Filter capacitance | 1 µF |
| $R_L$ | Load resistance | 21.4 Ω |

TABLE II
STEADY-STATE OPERATING POINT

| Symbol | Quantity | Value |
|---|---|---|
| $V_1$ | Dc input voltage | 20.0 V |
| $d_1$ | Inverter pulse density | 0.5 |
| $I_{L1r}$ | Real part of the transmitter resonant current phasor | 0.93 A |
| $I_{L2i}$ | Imaginary part of the receiver resonant current phasor | −1.16 A |
| $d_2$ | Rectifier pulse density | 0.5 |
| $V_2$ | Dc output voltage | 11.1 V |

The uncontrollable part (9) is linear and has a globally stable equilibrium point: $I_{L1i}(t) = I_{L2r}(t) = 0$. Therefore, only the controllable part (8) should be considered in the modeling, and it can be used as a simplified dynamical model under the tuned condition.

Moreover, a small-signal model can be derived from (8) to overcome the nonlinearity caused by the terms $d_2(t)V_2(t)$ and $d_2(t)I_{L2i}(t)$ as

$$\frac{d}{dt}\begin{bmatrix}\hat{I}_{L1r}(t)\\\hat{I}_{L2i}(t)\\\hat{V}_2(t)\end{bmatrix} = \begin{bmatrix}-\frac{R_1}{2L_1} & \frac{\omega_s M}{2L_1} & 0\\-\frac{\omega_s M}{2L_2} & -\frac{R_2}{2L_2} & \frac{\sqrt{2}d_2}{\pi L_2}\\0 & -\frac{2\sqrt{2}d_2}{\pi C_f} & -\frac{1}{R_L C_f}\end{bmatrix}\begin{bmatrix}\hat{I}_{L1r}(t)\\\hat{I}_{L2i}(t)\\\hat{V}_2(t)\end{bmatrix} + \begin{bmatrix}\frac{\sqrt{2}V_1}{\pi L_1} & 0\\0 & \frac{\sqrt{2}V_2}{\pi L_2}\\0 & -\frac{2\sqrt{2}I_{L2i}}{\pi C_f}\end{bmatrix}\begin{bmatrix}\hat{d}_1(t)\\\hat{d}_2(t)\end{bmatrix} \quad (10)$$

where $d_2$, $V_1$, $V_2$, and $I_{L2i}$ are the steady-state values.

## IV. EXPERIMENTAL TESTS

The WPT system and the modulation strategy described in [11] were used for the experiment. The circuit diagram of the system was as shown in Fig. 2. The specific parameters and the corresponding steady-state operating point are listed in TABLE I and TABLE II, respectively.

The first experiment tested the large-signal behaviors by measuring the time-domain responses of $i_{L1}(t)$, $i_{L2}(t)$, and $u_2(t)$ triggered by the step changes of $d_1(t)$ and $d_2(t)$ between 0.5 and 1, as shown in Fig. 3. The step up and down occurred at 0 ms and 0.5 ms, respectively. The transient waveforms of $i_{L1}(t)$, $i_{L2}(t)$, and $u_2(t)$ coincided with the model predicted envelopes given by $\sqrt{2}|I_{L1r}(t)|$, $\sqrt{2}|I_{L2i}(t)|$, and $V_2(t)$, which were derived from model (8).

The second experiment tested the small-signal behaviors by measuring the frequency-domain responses of the ripples on $V_2(t)$ stimulated by the small sine waves injected into $d_1(t)$ and $d_2(t)$ at the steady-state operating point. Fig. 4 shows the captured waveforms when the frequency of the injected sine wave was 10 kHz. At this frequency, the magnitude and phase of the output voltage ripples were compared to the sine wave, and the results were plotted in the Bode diagrams (see Fig. 5) as a pair of blue circles. By sweeping the frequency of the sine waves, more results were obtained and they all coincided with the transfer functions derived from model (10).

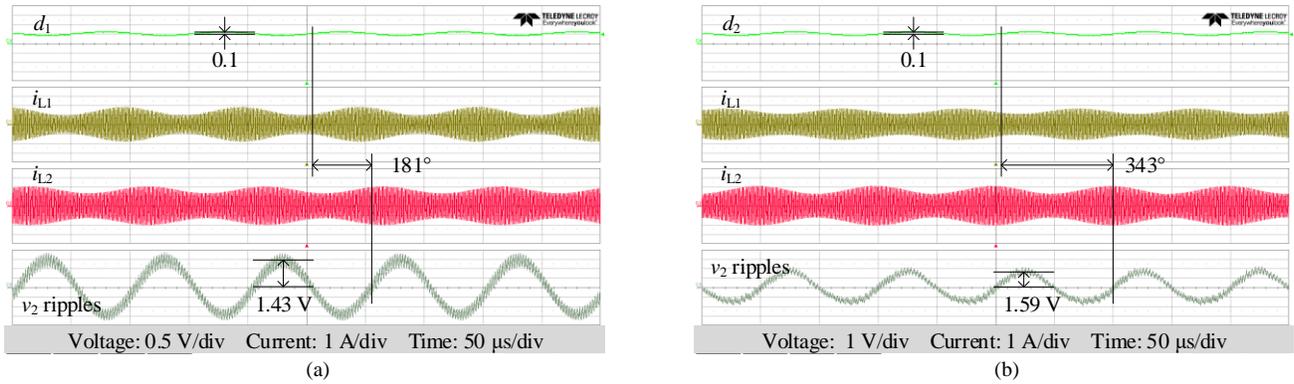

Fig. 4. Captured waveforms when injecting a 10 kHz sine wave into (a) $d_1$ and (b) $d_2$ at the steady-state operating point.

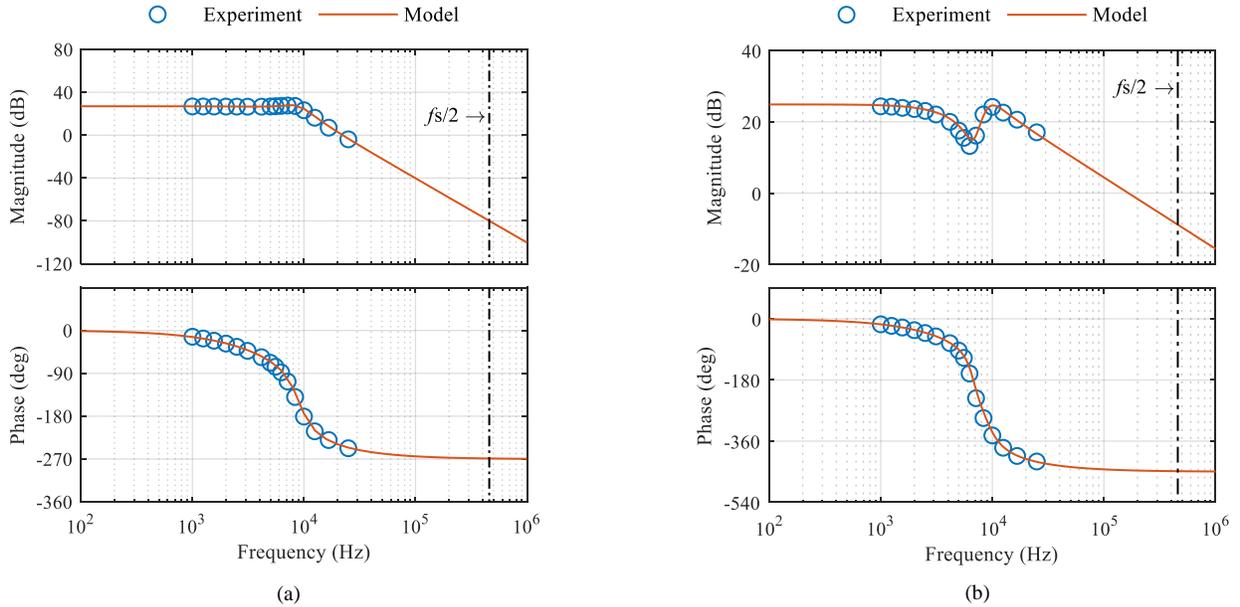

Fig. 5. Frequency-domain responses from the small signals of (a) $d_1$ and (b) $d_2$ to the small signal of $V_2$.

## V. CONCLUSION

This letter proposes a simplified dynamical model for the WPT systems under tuned condition (i.e., the rectifier input impedance is resistive and the two sides' resonant frequencies both equal the inverter fundamental switching frequency). The condition can be well satisfied by the recently proposed ZVS full-bridge PDM WPT system. Such a system contains 4 resonant elements and 1 filter capacitor, and in this letter, it is modeled by a simple 3$^{rd}$-order real-valued dynamical model. In contrast, existing dynamical models for the same resonant topology are usually 5$^{th}$- or 9$^{th}$-order models with much more complicated forms.

## REFERENCES

[1] H. Li, J. Fang, and Y. Tang, "Reduced-order dynamical models of tuned wireless power transfer systems," in *IEEE Int. Power Electron. Conf. ECCE Asia*, 2018, pp. 337-341.
[2] C. T. Rim and C. Mi, *Wireless power transfer for electric vehicles and mobile devices*. John Wiley & Sons, 2017.
[3] J. M. Miller, P. T. Jones, J. M. Li, and O. C. Onar, "ORNL experience and challenges facing dynamic wireless power charging of EV's," *IEEE Circ. and Syst. Mag.*, vol. 15, no. 2, pp. 40-53, 2015.
[4] H. Hao, G. A. Covic, and J. T. Boys, "An approximate dynamic model of LCL-T-based inductive power transfer power supplies," *IEEE Trans. Power Electron.*, vol. 29, no. 10, pp. 5554-5567, Oct 2014.
[5] Z. U. Zahid *et al.*, "Modeling and control of series-series compensated inductive power transfer system," *IEEE J. Emerg. Sel. Topics Power Electron.*, vol. 3, no. 1, pp. 111-123, Mar 2015.
[6] S. Lee, B. Choi, and C. T. Rim, "Dynamics characterization of the inductive power transfer system for online electric vehicles by Laplace phasor transform," *IEEE Trans. Power Electron.*, vol. 28, no. 12, pp. 5902-5909, Dec 2013.
[7] H. Li, J. Fang, and Y. Tang, "Dynamic phasor-based reduced order models of wireless power transfer systems," *IEEE Trans. Power Electron.*, 2019, Early Access.
[8] H. Li, J. Li, K. Wang, W. Chen, and X. Yang, "A maximum efficiency point tracking control scheme for wireless power transfer systems using magnetic resonant coupling," *IEEE Trans. Power Electron.*, vol. 30, no. 7, pp. 3998-4008, Jul 2015.
[9] *Qi wireless power specification*, 1.2.3, 2017.
[10] H. Li, K. Wang, J. Fang, and Y. Tang, "Pulse density modulated ZVS full-bridge converters for wireless power transfer systems," *IEEE Trans. Power Electron.*, vol. 34, no. 1, pp. 369-377, Jan 2019.
[11] H. Li, S. Chen, J. Fang, Y. Tang, and M. de Rooij, "A low-subharmonic, full-range, and rapid pulse density modulation strategy for ZVS full-bridge converters," *IEEE Trans. Power Electron.*, 2018, Early Access.